# AB-Space Engine*

**Alexander Bolonkin**

C&R, 1310 Avenue R, #F-6, Brooklyn, NY 11229, USA

T/F 718-339-4563, aBolonkin@juno.com, http://Bolonkin.narod.ru

## Abstract

On 4 January 2007 the author published the article "Wireless Transfer of Electricity in Outer Space" in http://arxiv.org wherein he offered and researched a new revolutionary method of transferring electric energy in space. In that same article, he offered a new engine which produces a large thrust without throwing away large amounts of reaction mass (unlike the conventional rocket engine).

In the current article, the author develops the theory of this kind of impulse engine and computes a sample project which shows the big possibilities opened by this new "AB-Space Engine". The AB-Space Engine gets the energy from ground-mounted power; a planet's electric station can transfer electricity up to 1000 millions (and more) of kilometers by plasma wires. Author shows that AB-Space Engine can produce thrust of 10 tons (and more). That can accelerate a space ship to some thousands of kilometers/second. AB-Space Engine has a staggering specific impulse owing to the very small mass expended. The AB-Space Engine reacts not by expulsion of its own mass (unlike rocket engine) but against the mass of its planet of origin (located perhaps a thousand of millions of kilometers away) through the magnetic field of its' plasma cable. For creating this plasma cable the AB-Space Engine spends only some kg of hydrogen.

----------------------------

## Introduction

**General information**.

A **rocket** is a vehicle, missile or aircraft which obtains thrust by the reaction to the ejection of fast moving fluid from within a rocket engine. Chemical rockets operate due to hot exhaust gas made from "propellant" acting against the inside of an expansion nozzle. This generates forces that both accelerate the gas to extremely high speed, as well as, since every action has an equal and opposite reaction, generating a large thrust on the rocket.

The history of rockets goes back to at least the 13th century, possibly earlier. By the 20th century it included human spaceflight to the Moon, and in the 21st century rockets have enabled commercial space tourism.

Rockets are used for fireworks and weaponry, as launch vehicles for artificial satellites, human spaceflight and exploration of other planets. While they are inefficient for low speed use, they are, compared to other propulsion systems, very lightweight, enormously powerful and can achieve extremely high speeds.

Chemical rockets contain a large amount of energy in an easily liberated form, and can be very dangerous, although careful design, testing, construction and use can minimise the risks.

A **rocket engine** is a jet engine that takes all its reaction mass ("*propellant*") from within tankage and forms it into a high speed jet, thereby obtaining thrust in accordance with Newton's third law. Rocket engines can be used for spacecraft propulsion as well as terrestrial uses, such as missiles. Most rocket engines are internal combustion engines, although non combusting forms also exist.

**Transfer of electricity into space**. The production, storage, and transference of large amounts of electric energy are an enormous problem for humanity, especially that of energy transfer in outer space (vacuum). Entire spheres of industry should be searching for, and badly need, revolutionary ideas. If in the production of energy, space launch and flight we have new ideas (see [1]-[17]), but we have not seen revolutionary ideas in transferring and storage energy except for reference [5].

However, if we solve the problem of transferring energy in outer space, then we solve many problems of manned and unmanned space flight. For example, spaceships can move long distances by using efficient electric engines, orbiting satellites can operate for unlimited time periods without falling prey to orbital decay and premature re-entry to Earth's atmosphere, communication satellites can transfer a strong signal directly to customers, the International Space Station's users can conduct many practical experiments and the global space industry can produce new materials. In the future, Moon and Mars outposts can better explore the celestial bodies on which they are placed at considerable expense [1].

Another important Earth mega-problem is efficient transfer of electric energy for long distances (intra-national, international, intercontinental). Nowadays, a lot of loss occurs from such energy transformation. The consumption of electric energy strongly depends on time (day or night), weather (hot or cold), and season (summer or winter). But an electric station can operate most efficiently in a permanent base-load generation regime. We need to transfer the energy long distance to any region that requires a supply in any given moment or to special pumped storage stations. One solution for this macro-problem is to transfer energy from Europe to the USA during nighttime in Europe and from the USA to Europe when it is night in the USA. Another solution is efficient energy storage, which allows people the option to save electric energy [1].

The storage of a big electric energy can help to solve the problem of cheap space launch. The problem of an acceleration of a spaceship can be solved by use of a new linear electrostatic engine suggested in [6]. However, the cheap cable space launch offered by author [5] requires the use of gigantic amounts of energy in short time period. (It is inevitable for any launch method because we must accelerate big masses to the very high speeds of 8 - 11 km/s). But it is impossible to turn off a whole state and connect an entire electric station to one customer. The offered electric energy storage can help solving this mega-problem for humanity [1]-[17].

**Railgun.**

The scientists used a railgun for high acceleration of small conducting body. A **railgun** is a form of gun that converts electrical energy (rather than the more conventional chemical energy from an explosive propellant) into projectile kinetic energy. It is not to be confused with a coilgun (Gauss gun). Rail guns use magnetic force to drive a projectile. Unlike gas pressure guns, rail guns are not limited by the speed of sound in a compressed gas, so they are capable of accelerating projectiles to extremely high speeds (many kilometers per second).

A wire carrying an electrical current, when in a magnetic field, experiences a force perpendicular to the direction of the current and the direction of the magnetic field.

In an electric motor, fixed magnets create a magnetic field, and a coil of wire is carried upon a shaft that is free to rotate. An electrical current flows through the coil causing it to experience a force due to the magnetic field. The wires of the coil are arranged such that all the forces on the wires make the shaft rotate, and so the motor runs.

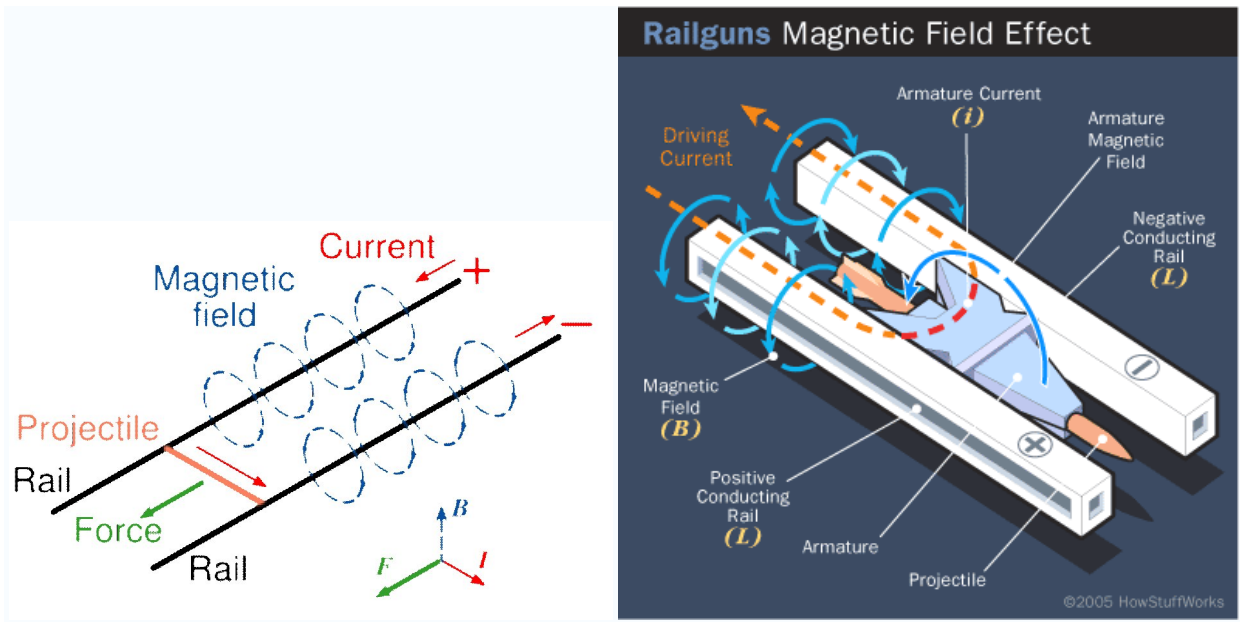

**Fig.1**. Schematic diagrams of a railgun.

A railgun consists of two **parallel** metal rails (hence the name) connected to an electrical **power supply**. When a conductive projectile is inserted between the rails (from the end connected to the power supply), it completes the circuit. Electrical current runs from the positive terminal of the power supply up the positive rail, across the projectile, and down the negative rail, back to the power supply (Fig.1).

This flow of current makes the railgun act like an **electromagnet**, creating a powerful magnetic field in the region of the rails up to the position of the projectile. In accordance with the **right-hand rule**, the created magnetic field circulates around each conductor. Since the current flows in opposite direction along each rail, the net magnetic field between the rails (**B**) is directed vertically. In combination with the current (**I**) flowing across the projectile, this produces a **Lorentz force** which accelerates the projectile along the rails. The projectile slides up the rails away from the end with the power supply.

If a very large power supply providing a million **amperes** or so of current is used, then the force on the projectile will be tremendous, and by the time it leaves the ends of the rails it can be travelling at many kilometres per second. 20 kilometers per second has been achieved with small projectiles explosively injected into the railgun. Although these speeds are theoretically possible, the heat generated from the propulsion of the object is enough to rapidly erode the rails. Such a railgun would require frequent replacement of the rails, or use a heat resistant material that would be conductive enough to produce the same effect.

The need for strong **conductive** materials with which to build the rails and projectiles; the rails need to survive the violence of an accelerating projectile, and heating due to the large currents and friction involved. The force exerted on the rails consists of a recoil force - equal and opposite to the force propelling the projectile, but along the length of the rails (which is their strongest axis) - and a sideways force caused by the rails being pushed by the magnetic field, just as the projectile is. The rails need to survive this without bending, and thus must be very securely mounted.

The power supply must be able to deliver large currents, with both **capacitors** and **compulsators** being common.

The rails need to withstand enormous repulsive forces during firing, and these forces will tend to push them apart and away from the projectile. As rail/projectile clearances increase, **arcing** develops, which causes rapid vaporization and extensive damage to the rail surfaces and the insulator surfaces. This limits most research railguns to one shot per service interval.

Some have speculated that there are fundamental limits to the exit velocity due to the inductance of the system, and particularly of the rails; but United States government has made significant progress in railgun design and has recently floated designs of a railgun that would be used on a naval vessel. The designs for the naval vessels, however, are limited by their required power usages for the magnets in the rail guns. This level of power is currently unattainable on a ship and reduces the usefulness of the concept for military purposes.

Massive amounts of heat are created by the electricity flowing through the rails, as well as the friction of the projectile leaving the device. This leads to three main problems: melting of equipment, safety of personnel, and detection by enemy forces. As briefly discussed above, the stresses involved in firing this sort of device require an extremely heat-resistant material. Otherwise the rails, barrel, and all equipment attached would melt or be irreparably damaged. Current railguns are not sufficiently powerful to create enough heat to damage anything; however the military is pushing for more and more powerful prototypes. The immense heat released in firing a railgun could potentially injure or even kill bystanders. The heat released would not only be dangerous, but easily detectable. While not visible to the naked eye, the heat signature would be unmistakable to infrared detectors. All of these problems can be solved by the invention of an effective cooling method.

Railguns are being pursued as weapons with projectiles that do not contain explosives, but are given extremely high velocities: 3500 m/s (11,500 ft/s) or more (for comparison, the M16 rifle has a muzzle speed of 930 m/s, or 3,000 ft/s), which would make their kinetic energy equal or superior to the energy yield of an explosive-filled shell of greater mass. This would allow more ammunition to be carried and eliminate the hazards of carrying explosives in a tank or naval weapons platform. Also, by firing at higher velocities railguns have greater range, less bullet drop and less wind drift, bypassing the inherent cost and physical limitations of conventional firearms - "*the limits of gas expansion prohibit launching an unassisted projectile to velocities greater than about 1.5 km/s and ranges of more than 50 miles [80 km] from a practical conventional gun system.*"

If it is even possible to apply the technology as a rapid-fire automatic weapon, a railgun would have further advantages in increased rate of fire. The feed mechanisms of a conventional firearm must move to accommodate the propellant charge as well as the ammunition round, while a railgun would only need to accommodate the projectile. Furthermore, a railgun would not have to extract a spent cartridge case from the breech, meaning that a fresh round could be cycled almost immediately after the previous round has been shot.

 **Tests.** Full-scale models have been built and fired, including a very successful 90 mm bore, 9 MJ (6.6 million foot-pounds) kinetic energy gun developed by DARPA, but they all suffer from extreme rail damage and need to be serviced after every shot. Rail and insulator ablation issues still need to be addressed before railguns can start to replace conventional weapons. Probably the most successful system was built by the UK's Defence Research Agency at Dundrennan Range in Kirkcudbright, Scotland. This system has now been operational for over 10 years at an associated flight range for internal, intermediate, external and terminal ballistics, and is the holder of several mass and velocity records.

The United States military is funding railgun experiments. At the University of Texas at Austin Institute for Advanced Technology, military railguns capable of delivering tungsten armor piercing bullets with kinetic energies of nine million joules have been developed. Nine mega-joules is enough energy to deliver 2 kg of projectile at 3 km/s - at that velocity a tungsten or other dense metal rod could penetrate a tank.
 The United States Naval Surface Warfare Center Dahlgren Division demonstrated an 8

mega-joule rail gun firing 3.2 kilogram (slightly more than 7 pounds) projectiles in October of 2006 as a prototype of a 64 mega-joule weapon to be deployed aboard Navy warships. Such weapons are expected to be powerful enough to do a little more damage than a BGM-109 Tomahawk missile at a fraction of the projectile cost.

Due to the very high muzzle velocity that can be attained with railguns, there is interest in using them to shoot down high-speed missiles.

## Offered Innovations and Brief Descriptions

  **1. Transfer of electricity by plasma cable**. The author offers a series of innovations that may solve the many macro-problems of transportation, energy and thrust in space. Below are some of them.

1) Transfer of electrical energy in outer space using a conductive cord from plasma. Author has solved the main problem - how to keep the plasma cord from dissipation, and in compressed form. He has developed the theory of space electric transference, made computations that show the possibility of realization for these ideas with existing technology. The electric energy may be transferred for hundreds millions of kilometers in space (including Moon and Mars) [1].

2) Method of construction of space electric lines and electric devices.

3) Method of utilization and tapping of the plasma cable electric energy.

4) Two methods of converting the electric energy to impulse (thrust) motion of a spacecraft (these two means are utilization of the magnetic field and of the kinetic energy of ions and electrons of the electric current).

5) Design of a triple electrostatic mirror (plasma reflector), which can reflect the plasma flow [1].

Below are some succinct descriptions of some constructions made possible by these revolutionary ideas.

  **1. Transferring electric energy in Space**. The electric source (generator, station) is connected to the distant location in space by two artificially generated rarefied plasma cables (Fig.1a). These cables can be created by a plasma beam [1, 8] sent from the Moon, Earth mounted super high tower, or from a space station in low Earth orbit, or a local base at the target location.  If the plasma beam is sent remotely from the Earth, a local reflector station is required at the target site or at a third location to turn the circuit back toward its' starting point and closure.

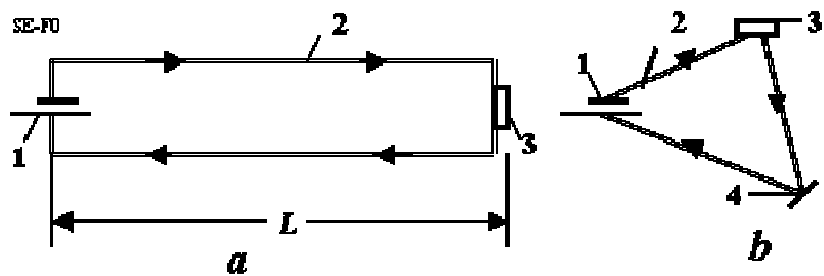

**Fig.2.** Long distance plasma transfer electric energy and thrust in outer space. **a** - plasma transfer with parallel plasma cable, **b** - plasma transfer with triangular (three-wire) plasma cable. *Notations*: 1 - current source (generator), 2 - plasma wire (cable), 3 - spaceship, orbital station or other energy destinations, 4 - plasma reflector located at planet, asteroid or space station.
The plasma cable may be ==also made from an ultra-cold (in radial direction) plasma.==

The plasma cable is self-supported in cable form by the magnetic field created by the electric current going through the plasma cable. The axial electric current produces an contracting magnetic pressure opposed to an expansive gas dynamic plasma pressure (the well-known theta-pinch effect)(Fig. 3). The plasma has a good conductivity (equal to that of silver and more) and the plasma cable can have a very big cross-section area (up to thousands of square meters cross-section). The plasma conductivity does not depend on its' density. That way the plasma cable has no large resistance although the length of plasma cable is hundreds of millions of kilometers. The needed minimum electric current is derived from parameters of a plasma cable researched in the theoretical section of this article.

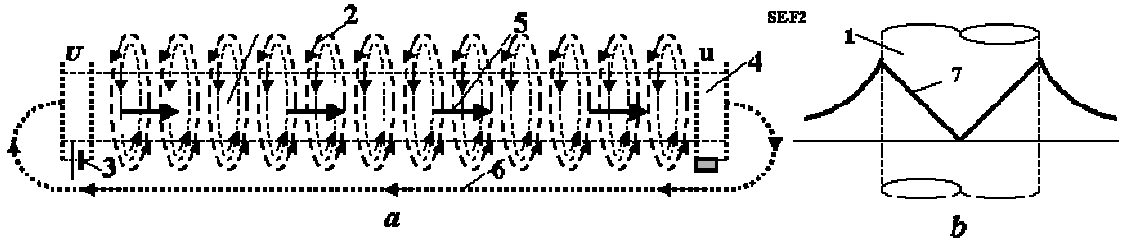

**Fig.3. *a*.** A plasma cable supported by its' own magnetic field, ***b***. Magnetic intensity into and out of plasma cable. *Notations*: 1 -plasma cable, 2 - compressing magnetic field, 3 - electric source, 4 - electric receiver, 5 - electric current, 6 - back plasma line; 7 – magnetic intensity into and out of plasma cable.

The parallel cables having opposed currents repel each other (Fig.2a)(by magnetic force). This force may be balanced by attractive electric force if we charge the cables by electric charges (see theoretical section). They also can be separated by a special plasma reflector as it shown in figs. 2b. The electric line can be created and exist independently. The spaceship connects to this line at a suitable point.  By altering the diameter and direction of the plasma cable we can supply energy to a spacecraft. Though we must supply energy to accelerate the spacecraft we can also regenerate energy by braking it.  At any time the spaceship can disconnect from the line and can exist without line support (propulsion, electricity, etc). The apparatus can hook up to or disconnect from the plasma cable at will. But breaking (loss of continuity) of the plasma cable itself destroys the plasma cable line to the remote location! We must have additional (parallel) plasma lines and apparatus must disconnect from a damaged or occulted (for example on the far side of a remote planet) plasma line and connect to another line to keep the connection in existence. The same situation is true in a conventional electric net. The apparatus can also restore the damaged part of plasma line by own injected plasma, but the time for repairing is limited (by tens of minutes or some hours). The original station can also to send the plasma beam which connects the ends of damaged part of the line.

The electric tension (voltage) in a plasma cable is between two ends (for example, as cathode-anode) of the conductor in the issuing electric station (electric generator) [1,8,9]. The plasma cable current has two flows:
Electron (negative) flow and opposed ions (positive) flow in the same cable. These flows create an electric current. (In metal we have only electron flow, in liquid electrolytes we have ions flowing).
   The author offers methods (for extraction and inserting) of energy from the plasma electric cable (Fig.4) by customer (spacecraft, other energy destination or end user).

The double net can accelerate the charged particles and insert energy into plasma cable (fig, 4a) or brakes charged particles and extract energy from electric current (fig. 4b). In the first case the two nets create the straight electrostatic field, in the second case the two nets create the opposed electrostatic field in plasma cable (resistance in the electric cable [1, 8, 9]) (figs.2, 4c). This apparatus resistance utilizes the electric energy for the spaceship or space station. In the second case the charged particles may be collected into a set of thin films and emit (after utilization in apparatus) back into continued plasma cable (see [1,8,9]).

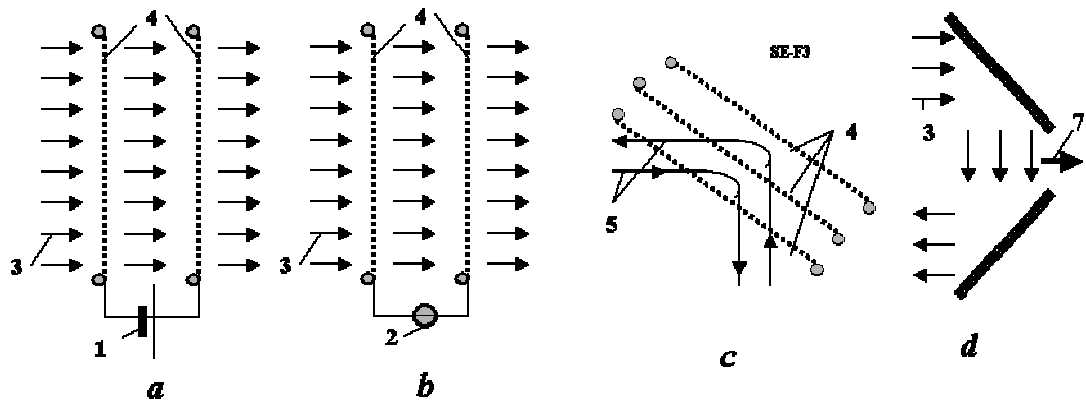

**Fig.4.** Getting and inserting in (off) plasma cable the energy and turning of plasma cable. *a* – inserting electric energy into plasma cable by means of two thin conducting nets or films; *b* - getting the energy from plasma cable by means of two thin conducting nets or films; *c* – offered triple net plasma reflector; *d* – double triple net plasma reflectors - the simplest AB thruster. *Notations*: 1 – spaceship or space station, 2 – receiver of energy, 3 - plasma cable, 4 - electrostatic nets, 5 – two opposed flows of charged plasma particles (negative and positive: electron and ions), 7 – thrust of AB-Space Engine.

Fig. 4c presents the plasma beam reflector [1,8,9]. That has three charged nets. The first and second nets reflect (for example) positive particles, the second and third nets reflected the particles having an opposed charge.

Fig.5 shows the different design the plasma cable in space.

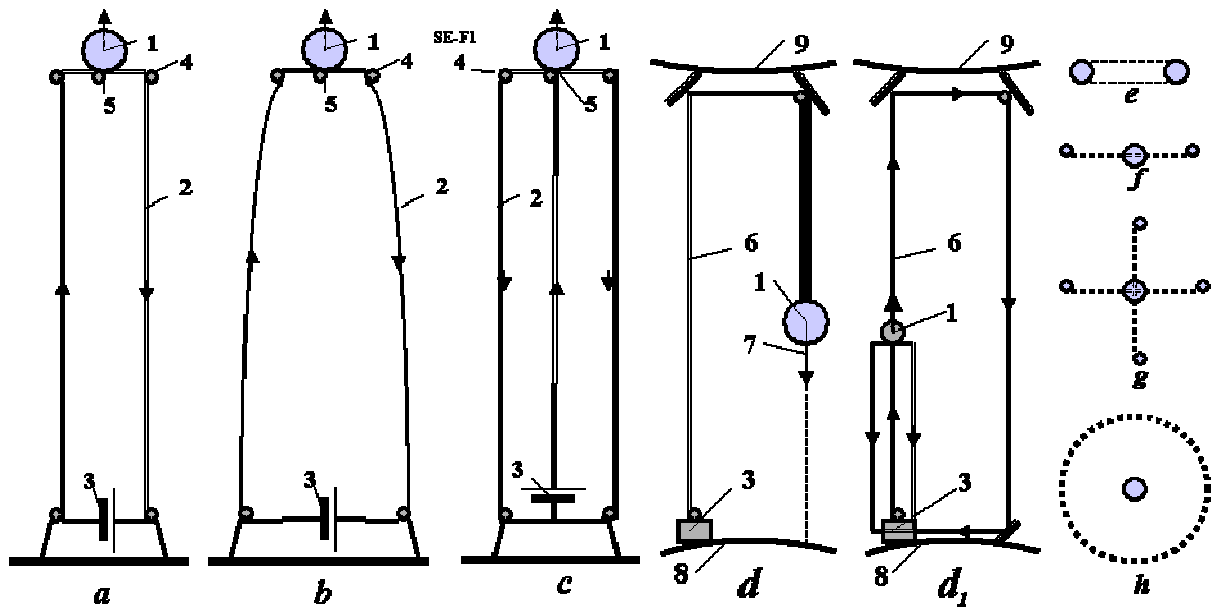

**Fig.5.** Transfer electricity and thrust by AB-Space Engine: *a*. Two plasma parallel cables; *b*. Curved cable; *c.* Plasma multi-cables; *d*. Transfer of back thrust through planet or asteroid; *d*₁. Using of ready plasma line; *e – h*. Forms of straight and back plasma cables (cross-sections of cables). *Notations*: 1 – Space ship; 2 – plasma cable; 3 – source electricity; 4 - plasma injector; 5 – user of energy; 6 – double plasma line; 7 - thrust; 8 – Earth; 9 – planet or asteroid.

Fig.5a shows two plasma parallel cables. Fig.5b two shows plasma parallel cables of a curved form of line. Fig.5c presents three plasma parallel cables, one to space ship and two for back (return) current. Fig.5d shows the transfer of the reverse impulse (or braking) thrust to space ship through planet or asteroid. Figs.5e-h shows the different forms of the straight and back plasma cables (cross-section of cables).

**2. AB-Space Engine.** The offered simplest AB-Space Engine is shown in Fig. 4d and more details in Fig. 6a. That includes two new triple electrostatic reflectors 2 which turn the plasma cables' (flow), (electric current 3) in back direction. The engine may also contain (optional) the plasma injectors 5 and electric generator (user) 4.

As feed material for the plasma may be used hydrogen gas, as plasma reflector may be used three conductivity nets connected to voltage sources, as generator - the double conductivity nets located into plasma flow and connected to voltage sources or users.

The other design of AB-Space Engine is shown in fig.6b. Here the central plasma flow divides in two side flows which go back to the electric station.

The AB-Space Engine works as follows. The electric current (voltage) produced by electric station (that may be located far from AB-Space Engine, for example, in orbit around the Earth or mounted on the Moon, Phobos or another space body) transfers by plasma cable to the AB-Space Engine. The power of the electric current in the plasma produces the power plasma flow of electrons and ions. The engine turns back the plasma flow (electric current) and returns it to the source electric station by the other plasma cable. The magnetic and centrifugal forces appear at the point of turning from outgoing to ingoing plasma paths place and create the thrust which can be used for movement (acceleration, braking) the space apparatus (or conventional vehicle or projectile).

Long-time readers of proposed space drive papers may suspect something fishy here. Don't worry: The AB-Space Engine doesn't violate Newton's third law of action and reaction. The AB-Space Engine reacts against the (planet or station mounted) electric station which may be located hundreds of millions of kilometers away! No other engine has the same capability.

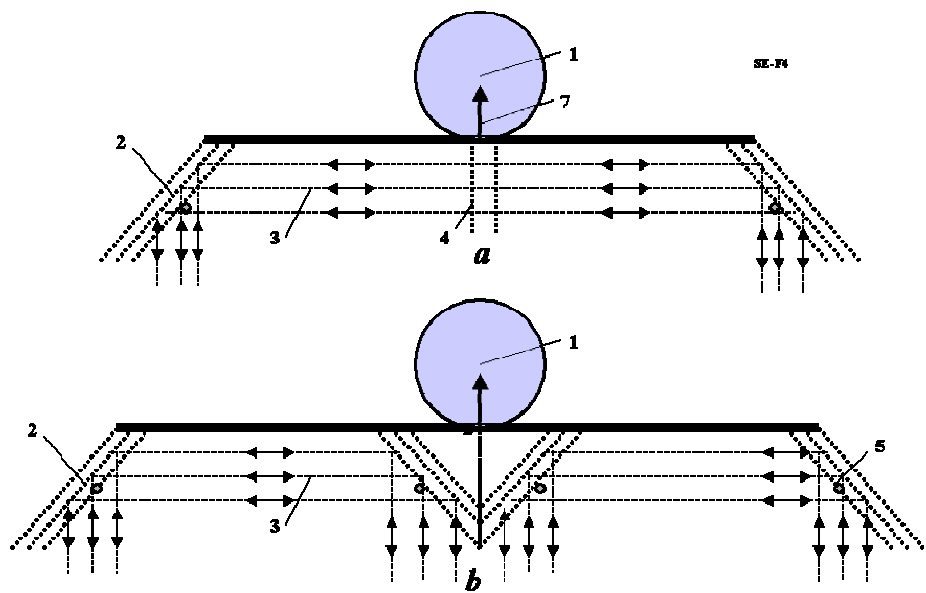

**Fig.6.** Some versions of AB-Space Engine (thruster). *a*. two cable AB-Space Engine; *b*. Three cable AB-Space Engine. *Notations*; 1 – space ship; 2 – offered special (three nets) electrostatic reflector; 3 – plasma cable; 4 – receiver or source of energy; 5 – injector of plasma, 7 – thrust.

Your attention is also directed to the following differences between a railgun and an AB-Space Engine:

1) The railgun uses SOLID physical rails for delivery the electric current to conductivity projectile. These are easily damaged by huge electric current. The AB-Space Engine uses flexible plasma cables which can self-repair.

2) The railgun uses the rails which are of fixed construction (unalterable) and a spacecraft so launched can move solely in the rail direction. The AB-Space Engine creates the plasma cable in the course of apparatus movement and can select and change the apparatus' future direction.

3) Even a theoretical railgun girdling the globe of the Moon in vacuum (for star probe launch)

would have a possible length of only some kilometers (as any solid construction). The plasma electric line (used by AB-Space Engine) can have a length (an acceleration path) of millions of kilometers (and thus may someday power manned craft on missions to near interplanetary space).

## Theory of AB-Engine, Estimations and Computations

### 1. General Theory of AB-Engine and Transfer Electricity in Space.

*The magnetic intensity and magnetic pressure* of an electric current reaches a maximum upon the surface of a plasma cable. Let us attempt to equate plasma gas pressure to a magnetic pressure and find the requested equilibrium electric current for a given (same) temperature of electrons and ions

$$P_g = 2nkT_k, \quad P_m = \frac{\mu_0 H^2}{2}, \quad H = \frac{I}{2\pi r},$$

$$P_m = P_g, \quad I = 4\pi r \left( \frac{k n T_r}{\mu_0} \right)^{0.5}, \quad T_k = \frac{m_e u_r^2}{2k}, \tag{1}$$

where $P_g$ is plasma gas pressure, N/m$^2$; $P_m$ is magnetic pressure, N/m$^2$; $n$ is plasma density (number of electron equals number of ions: $n = n_e = n_i$), 1/m$^3$; $k = 1.38 \times 10^{-23}$ is Boltzmann coefficient, J/K; $\mu_0 = 4\pi 10^{-7}$ is magnetic constant, H/m; $H$ is magnetic intensity, A/m; $I$ is electric current, A; $r$ is radius of plasma cable, m; $T_r$ is plasma temperature in radial direction of plasma cable, K; $m_e = 9.11 \times 10^{-31}$ is electron mass, kg; $u_r$ is average electron speed in radial direction of plasma cable, m/s.

*Minimal Electric current.* From (1) we receive relation between a minimal electric current $I_{min}$, gas density $n$ and the radial temperature of electrons

$$I_m = 4\pi r \left( \frac{k n T_r}{\mu_0} \right)^{0.5} \approx 4.16 \times 10^{-8} r \sqrt{n T_r},$$

$$j_m = \frac{I}{\pi r^2} = 4 \left( \frac{k}{\mu_0} \right)^{0.5} \frac{\sqrt{n T_r}}{r} \approx 1.33 \cdot 10^{-8} \frac{\sqrt{n T_r}}{r}, \tag{2}$$

where $I_m$ is minimal electric current, A; $j_m$ is density of electric current, A/m$^2$; $\pi r^2 = S$ is the cross-section area of plasma cable, m$^2$.

Assume the temperature (energy) of electrons equals temperature (energy) of ions. Let us to write well-known relations

$$j = en(u_i + u_e), \quad \frac{m_i u_i^2}{2} = \frac{m_e u_e^2}{2}, \tag{3}$$

where $e = 1.6 \times 10^{-19}$ C is charge of electron, C; $m_e = 9.11 \times 10^{-31}$ kg is mass of electron, kg; $m_i$ is mass of ion, kg (for H$_2$ $m_i = 2 \times 1.67 \times 10^{-27}$ kg); $u_i$, $u_e$ is speeds of ions and electrons respectively **along** cable axis produced by electric intensity (electric generator), m/s.

The computation $j$ by Eq. (2) is presented in Figure 7.

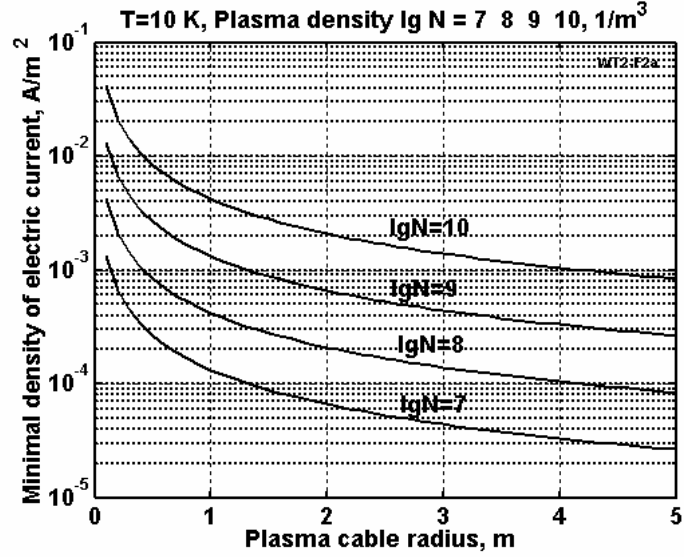

**Fig. 7**. Minimal density of electric current in plasma cable for radial plasma temperature $10^o$K.

From (3) we receive axial speeds of ions and electrons produced by electric intensity (electric generator)

$$u_e = \frac{j}{en(1+\sqrt{m_e/m_i})}, \quad u_i = \frac{j}{en(1+\sqrt{m_i/m_e})} \quad . \qquad (4)$$

or

$$u_e \approx 6{,}15 \cdot 10^{18}\, j/n \quad for\ H_2 \quad u_i \approx 10^{17}\, j/n, \quad u_e >> u_i, \quad u = u_e + u_i \approx u_e . \ (4)'$$
$$or \quad u \approx j/en .$$

Under electric intensity the electrons and ions have opposed speeds along cable axis. The computation of electron speed produced by minimal electric current is presented in Fig.8.

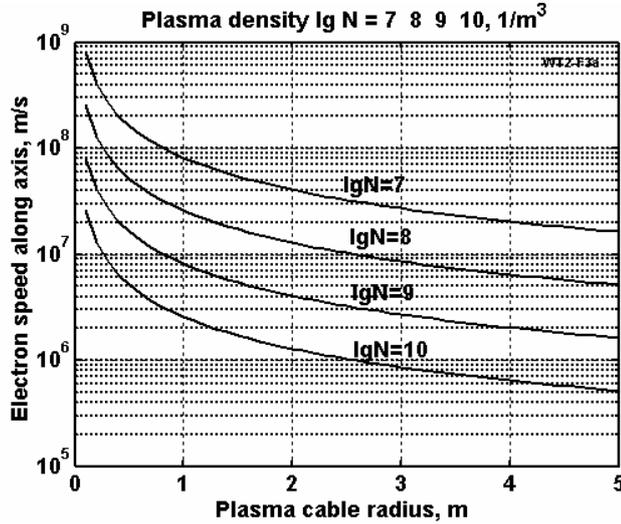

**Fig. 8**. The electron speed produced by the electric current of the minimal current density versus plasma cable radius. The ions (H$_2$) speed is less 61.5 times and opposed the electron speed.

*Temperature along plasma cable axis* induces by minimal electric voltage is

$$T_k = \frac{m_e u^2}{2k} \approx 3.3 \cdot 10^{-8} u^2 \quad [K], \quad T = \frac{k}{e} T_k \approx 2.71 \cdot 10^{-12} u^2 \quad [eV] , \quad (5)$$

where $T_k$ is induced temperature in K; $T$ is this temperature along cable axis in eV. Computation is shown in Fig.9.

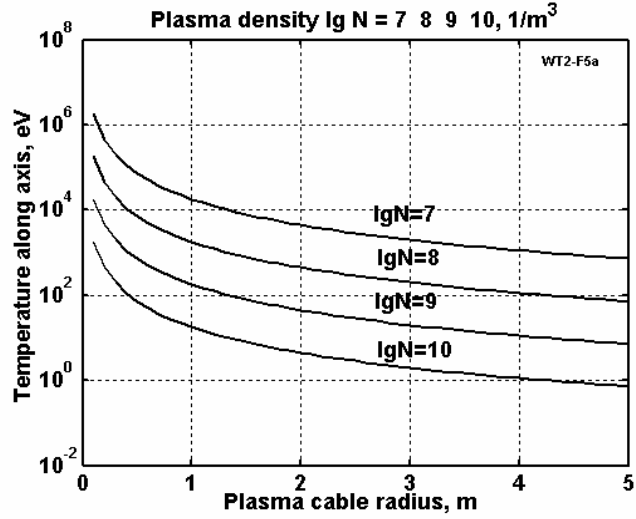

**Fig. 9**. The temperature of electron and ions (H₂) (in eV) produced by the electric current in the minimal current density versus the plasma cable radius and different plasma density. It is assumed the ions (H₂) temperature equals the electron temperature.

*Specific Spitzer plasma resistance* (the so-called Spitzer Conductivity) and typical resistance of a plasma cable can be computed by equations:

$$\rho = \eta_\perp = 1.03 \times 10^{-4} Z \ln \Lambda T^{-3/2} \quad \Omega \cdot m, \quad R = \rho L / S, \qquad (6)$$

where $\rho$ is specific plasma resistance, $\Omega \cdot m$; $Z$ is ion charge state, $\ln \Lambda \approx 5 \div 15 \approx 10$ is the Coulomb logarithm; $T = T_k k/e = 0.87 \times 10^{-4} T_k$ is plasma temperature along cable axis in eV; $e = 1.6 \times 10^{-19}$ is electron charge, C; $R$ is electric resistance of plasma cable, $\Omega$; $L$ is plasma cable length, m; $S$ is the cross-section area of the plasma cable , m².

The computation of the specific resistance of a plasma cable for minimal electric current is presented in Figure 10.

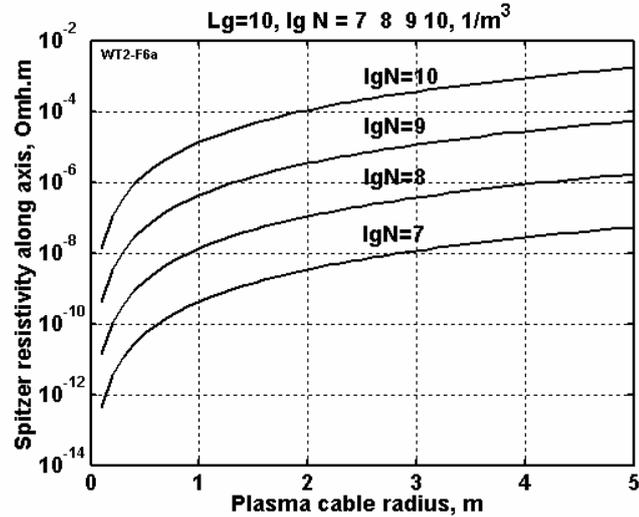

**Fig. 10.** Specific (Spitzer) plasma resistance $\Omega \cdot m$ of equilibrium plasma cable for the minimal electric current versus cable radius and different plasma density. Coulomb logarithm equals 10.

*The requested minimum voltage, power*, transmitter power and coefficient of electric efficiency are:

$$U_m = IR, \quad W_m = IU_m, \quad U = U_m + \Delta U, \quad W = IU, \quad \eta = 1 - W_m / W = 1 - U_m / U, \qquad (7)$$

where $U_m$, $W_m$ are requested minimal voltage, [V], and power, [W], respectively; $U$ is used voltage, V; $\Delta U$ is electric voltage over minimum voltage, V; $W$ is used electric power, W; $\eta$ is coefficient efficiency of the electric line. If $\Delta U >> U_m$ the coefficient efficiency closed to 1.

Computation of loss voltage and power into plasma cable having length 100 million km is in Figs. 11-12.

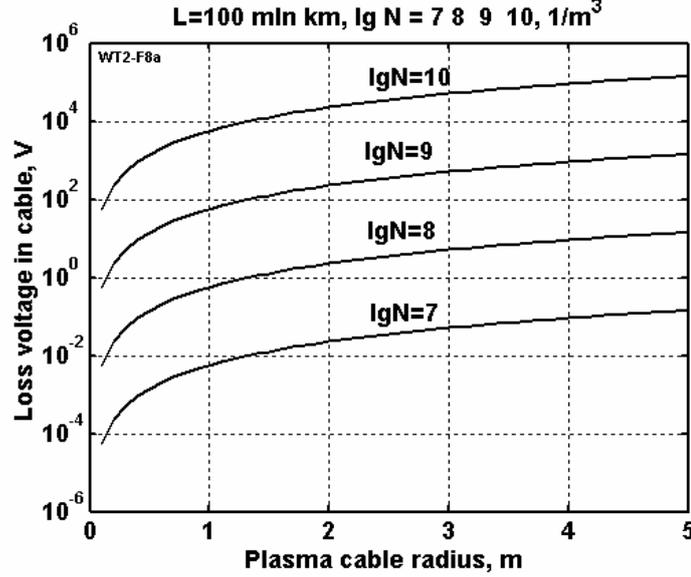

**Fig. 11**. Loss voltage in plasma cable of 100 millions km length via cable radius for the minimal electric current and different plasma density.

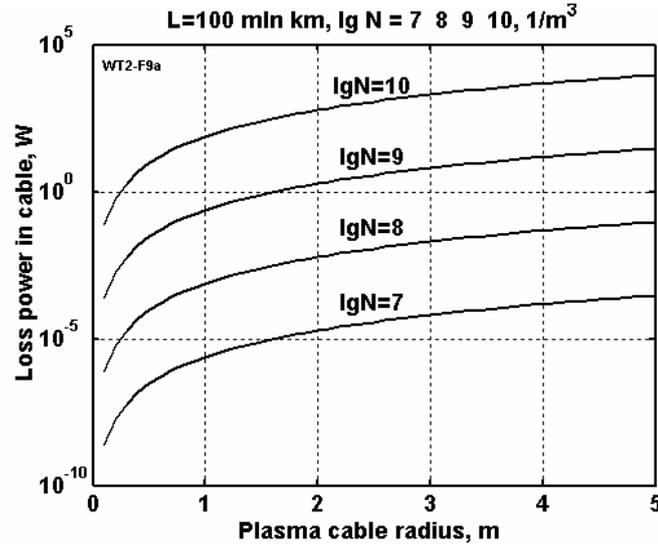

**Fig. 12**. Loss power in plasma cable of 100 millions km length via cable radius for the minimal electric current and different plasma density.

*The mass M* [kg] *of one plasma* cable is

$$M = \pi r^2 n m_i L, \qquad (8)$$

where $m_i$ is ion mass of plasma, kg; $L$ is length of plasma cable, m.

The mass of plasma cable is very small, about some grams for 100 millions km for $n < 10^{10}$/m³. The mass of a plasma cable is close to zero for any practical case when R < 5 m.

*The force acting in a particle* (proton) moved in electric and magnetic fields may be computed by the equations:

$$\overline{F}_1 = \frac{m_i v^2}{r}, \quad \overline{F}_2 = e\overline{v}\,\overline{B}, \quad \overline{F}_3 = \frac{eQ_0}{4\pi\varepsilon_0 R^2}, \quad \overline{F}_4 = \gamma\frac{m_1 m_2}{R^2}, \quad \overline{F}_4 = g m_i \qquad (9)$$

where $F_1$, $F_2$, $F_3$, $F_4$ are centrifugal, Lorenz, electrostatic, and gravitational forces respectively (all vectors), N; $m_p = 1.67{\times}10^{-27}$ kg mass of proton (or ion $m_i$); $v$ - speed of particle, m/s; $e$ - electron (proton) charge; $B$ - total magnetic induction (magnetic field strength), T; $Q_0$ - charge of central body, C; $\varepsilon_0 = 8.85{\times}10^{-12}$ F/m - electric constant; $m_1$, $m_2$ are mass of bodies (central and particle), kg; $\gamma$ – gravitational constant (for Earth $\gamma = 6.67 \times 10^{-11}$ m$^3$/kg·s$^2$, $g_o = 9.81$ m/s$^2$; for Sun $g_o = 274$ m/s$^2$); $r$ – radius curve, m; $R$ – distance between charges (gravitational bodies), m.

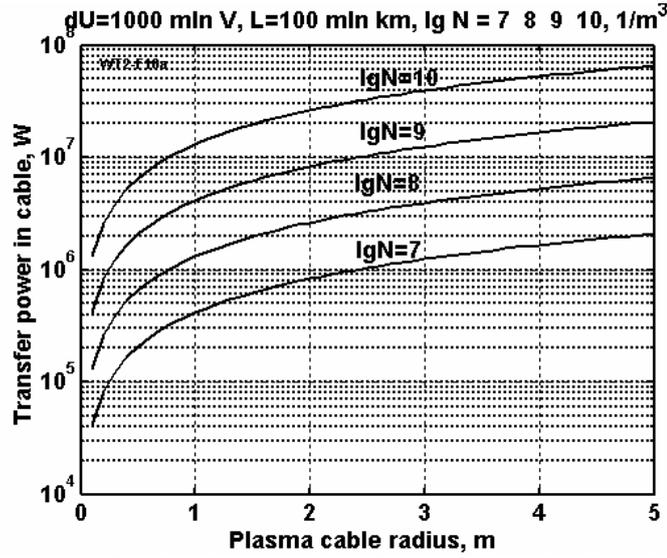

**Fig. 13**. Electric power transfers by plasma cable of 100 millions km length via cable radius for the minimal electric current, over voltage $10^9$ V and the different plasma density. Coefficient efficiency is about 0.9999.

The equilibrium condition is:

$$\sum_i F_i = 0 \, .$$

*Magnetic Pressure (magnetic thrust) from the Plasma Cable*. The plasma exerts a pressure within the plasma cable. This pressure is small, but the cable can has a large diameter (up 200 m or more) and this pressure acting over a long time can accelerate or brake a space apparatus with no reaction mass. This magnetic pressure $P$ [N/m$^2$] from only one cable can be computed by equations:

$$P_m = \frac{\mu_0 H^2}{2}, \quad H = \frac{I}{2\pi r}, \quad P = \frac{1}{2} 2 P_m S = \frac{\mu_0}{4\pi} I^2 \, . \tag{10}$$

*Estimation*. For $I = 10^4$ A, the magnetic pressure equals 10 N; for $I = 10^5$ A, it equals 1000 N; for $I = 10^6$ A, the thrust of one cable is $P = 10^5$ N = 10 tons.

That is magnetic thrust from one cable. The AB-Space Engine has two cables (incoming and out coming), that means the magnetic thrust from two cable will be (at minimum) two times more. If we compute the horizontal part of plasma cable which is pressed by outer plasma magnetic field the full thrust is:

$$P = \frac{\mu_0}{2\pi} I^2 \ln\left|\left(\frac{d-r}{r}\right)^2\right|, \tag{11}$$

where $d$ is distance between centers of the incoming and out coming plasma cables.

*Electric (kinetic) Pressure from the Plasma Cable*. The high speed electrons and ions of electric current within plasma cable have kinetic energy. This energy produces kinetic (electric) pressure when space ship or final station uses the electric energy. Let as to estimate the electric pressure.

Specific (kinetic) energy of electric current into plasma cable is

$$E = 0.5n(m_e u_e^2 + m_i u_i^2), \quad [\text{W/m}^3] \tag{12}$$

Substitute Eqs.(4) in (12) we have

$$E = \frac{j^2}{2ne^2}\left[\frac{m_e}{(1+\sqrt{m_e/m_i})^2} + \frac{m_i}{(1+\sqrt{m_i/m_e})^2}\right] \approx \frac{j^2}{2ne^2}[m_e + m_e] = \frac{m_e}{e^2}\frac{j^2}{n}. \quad (13)$$

But specific energy equals the specific pressure $P_e = E_s$ [N/m$^2$].

$$P_e = \frac{m_e}{e^2}\frac{j^2}{n} \approx 3.36 \times 10^7 \frac{j^2}{n}. \quad (14)$$

*Estimation*: For $j = 100$ A/m$^2$, $n = 10^{10}$ 1/m$^3$ we get $P = 35.6$ N/m$^2$ (for comparison, 1 Pa = 1 N/m$^2$ = 10$^{-5}$ bar).

Full kinetic energy of charged particles the plasma cable is $E = P_e sL$ [J], where $s$ is cross-section area of plasma cable [m$^2$], $L$ is length of plasma cable [m].

*Additional Power from a Space Apparatus' Motion.*
This power is:

$$W = PV, \quad (15)$$

where $V$ is apparatus speed, m/s.

*Estimation*. For $V = 11$ km/s, $P = 10^{-3}$ N, this power equals 11 W, for $P = 1$ N the power equals 11000 Watts. We spend this power when space apparatus moves away from the energy source ('launch point') and receive it when apparatus approaches to the energy station. ('landing site')

*Track Length of Plasma Electrons and Ions.*
The track length $L$ and the track time $\tau$ of particles is

$$L = \upsilon_T/\nu, \quad \tau = 1/\nu, \quad (16)$$

where $\upsilon_T$ is particle velocity, cm/s; $\nu$ is particle collision rate, 1/s.
The electron, ion, and electron-ion **thermal** collision rate are respectively:

$$\nu_e = 2.91 \times 10^{-6} n_e \ln\Lambda T_e^{-3/2} \quad s^{-1}$$
$$\nu_i = 4.80 \times 10^{-8} Z^4 \mu^{-1/2} n_i \ln\Lambda T_i^{-3/2} \quad s^{-1}, \quad (17)$$
$$\nu_{ei} = 4.4 \times 10^{-6} n_i \lg\Lambda T^{-3/2}.$$

where $Z$ is ion charge state, $\ln\Lambda \approx 5 \div 15 \approx 10$ is Coulomb logarithm, $\mu = m_i/m_p$ is relative mass of ion; $m_p = 1.67 \times 10^{-27}$ is mass of proton, kg; $n$ is density of electrons and ions respectively, 1/cm$^3$; $T$ is temperature of electron and ion respectively, eV.
Electron and ion **thermal** velocity are respectively:

$$\upsilon_{Te} = (kT_e/m_e)^{1/2} = 4.19 \times 10^7 T_e^{1/2} \quad \text{cm/s}$$
$$\upsilon_{Ti} = (kT_i/m_i)^{1/2} = 9.79 \times 10^7 \mu^{-1/2} T_i^{1/2} \quad \text{cm/s}, \quad (18)$$

Substitute equations (12)-(13) in (11) we receive the length of electron and ion tracks:

$$L_e = 1.44 \times 10^{13} T_e^2/n_e \ln\Lambda \quad \text{cm},$$
$$L_i = 2.04 \times 10^{13} T_e^2/Z^4 n_e \ln\Lambda \quad \text{cm}, \quad (19)$$
$$L_{ei} = 0.95 \times 10^{13} T_e^2/n_e \ln\Lambda \quad \text{cm}.$$

*Estimation*. For electron having $n = 10^5$ 1/cm$^3$, $T = 100$ eV, $\ln\Lambda \approx 10$ we get $L = 2 \times 10^6$ km, $\tau \approx 300$ s.
That means the plasma electrons have very few collisions, small dispersion, (in our case) and it can have different average ELECTRON (relative to ion) temperature along the cable axis and perpendicular cable axis. It is not a surprise because the plasma can have different average temperatures of electron and ions. That also means that our assumption about the terminal and current electron velocities being the same is very limited and the parameters of a plasma electric system will often be better, than in our computation. The *plasma in our system may be very cool in a radial direction and simultaneously very hot in the axial direction.* That decreases the electric current needed for plasma compression and allows a transfer of the plasma beam, energy, and thrust to a great distance.

*Magnetic force between two parallel cables*. This force is

$$F_m = -\mu_0 \frac{i_1 i_2}{2\pi d} L \ , \qquad (20)$$

where $F_m$ is magnetic force, N (the force is repeal when currents are opposed, and attractive when currents have same directions); $\mu_0 = 4\pi 10^{-7}$ is permeability constant, H/m; $i$ is electric current in the 1-st and 2-nd cable respectively, A; $d$ is distance between center of cables, m; $L$ is length of cables, m.

This force for two cable line (fig.5e) having current $I = 10^5$ A, distance $d = 1000$ km equals $F_m = 2$ N/km. But force decreases if we use multi-cable system: for three cables (Fig.5f)(3/8 $F_m$); for 5 cables (Fig.5g)(5/32 $F_m$); for multi-cables (Fig.5h)($F_m = 0$).

*Electrostatic force between two parallel cables*. This force is

$$F_e = k \frac{2\tau_1 \tau_2}{d} L \ , \qquad (21)$$

where $F_e$ is electrostatic force, N (the force is attractive when charges is different and repeal when charges are same); $k = 1/4\pi\varepsilon_0 = 9\times10^9$ electrostatic constant, Nm$^2$/C$^2$; $\tau$ is linear charge of the 1-st and 2-nd cable, C/m; $d$ is distance between cables, m; $L$ is length of cables, m.

Electrostatic force is attractive force for opposed charges. This force may be used for balance the electromagnetic force. From $F_m = F_e$, (20) = (21) we get for two line cable system

$$\tau = \frac{1}{2}\sqrt{\frac{\mu_0}{\pi k}} I, \quad \Delta U = 2k\tau \int_r^d \frac{dR}{R} = \sqrt{\frac{\mu_0 k}{\pi}} I \ln\left(\frac{d}{r}\right) = 60 I \ln\left(\frac{d}{r}\right), \quad \Delta P = \Delta U \cdot I \ , \quad (22)$$

where $r$ is plasma cable radius, m. Example: for $I = 10^4$ A, $d/r = 10$ we have $\Delta U = 1.38\times10^6$ V,

The linear charge appears on cable when is voltage between cables. The other way of balance is cable design in Fig.5h.

*Electric capacity two parallel cables* is

$$C = \frac{\pi\varepsilon_0}{\ln(d/r)} L \ , \qquad (23)$$

where $C$ is electric capacity, F; $\varepsilon_0 = 8.85\times10^{-12}$ is electrostatic constant, F/m; $r$ is cable radius, m;

*Energy of two parallel cables as electric condenser* is

$$E = \frac{1}{2} CU^2 = \frac{1}{2} qU = \frac{1}{2}\frac{q^2}{C} \ ,$$

where $E$ is energy in condenser, J; $U$ is electric voltage, V; $q$ is electric charge, C. Example: for $d = 100$ km, $r = 10$ m, the electric capacity is $C = 0.05$ F/one million km, the energy is $E = 2.5\times10^6$ J/one million km.

*Inductance of two parallel cables* is

$$L_i = \frac{\mu_0}{\pi}\left(\frac{1}{2} + \ln\frac{d}{r}\right) L \ , \qquad (24)$$

where $L_i$ is inductance, H.

*Inductance energy of two parallel cables* is

$$E = L_i \frac{I^2}{2} \ , \qquad (25)$$

where $E$ is energy in a closed-loop contour, J. Example: for $d = 100$ km, $r = 10$ m, the inductance is $L_i = 3.9\times10^3$ H/one million km, the energy is $E = 1.94\times10^{15}$ J/one million km. This energy is high and the starting station (where the plasma cables originate) spends a lot of energy for creating the magnetic field.

*Change electric current* in closed-loop contour is

$$I = I_0 \exp\left(-\frac{t}{T}\right), \quad \text{where} \quad T = \frac{L_i}{R} \ , \qquad (26)$$

where $R$ is electric resistance of closed-loop electric contour, Ohm. $T$ is time decreasing current by factor $e = 2.71$ times. Example: for two lines cable the length $L = 100$ millions of km and the electric resistance $R = 10^{-3}$ Ohm (see project below), $T \approx 3.9 \times 10^3 / 10^{-3} = 3.9 \times 10^6$ sec = 45 days. That means our electric line is a large storage reservoir of energy.

*'Virtual' Specific Impulse of AB-Space Engine.* Specific impulse of rocket engine is ratio of an engine thrust to fuel consumption per second. It is difficult to speak about specific impulse of AB-engine because AB-engine doesn't spend fuel for the thrust, but it does expend matter (for example hydrogen) for creating the plasma cables. That way we can take ratio of the thrust to mass expenditure per second for produce of new cable.

This 'virtual' specific impulse of the AB-Space Engine is

$$I_y = \frac{P}{m_s}, \quad P = \frac{\mu_0}{2\pi} I^2, \quad m_s = 2snm_iV, \quad I_y = \frac{\mu_0}{3\pi m_i} \frac{I^2}{snV}. \quad \text{For} \quad H_2 \quad I_y = 6 \cdot 10^{19} \frac{I^2}{snV} \ , \quad (27)$$

where $I_y$ is specific impulse of AB-Space Engine, m/s; $P$ is magnetic thrust , N; $s = \pi r^2$ is cross section area of plasma cable, m$^2$; $m_i = 2 \times 1.67 \times 10^{-27}$ is mass of one molecule of hydrogen, kg; $V$ is speed of apparatus, m/s.

*Estimation.* Let us take the $I = 10^6$ A, $s = 10^2$ m$^2$. $n = 10^{14}$ /m$^3$, $V = 10^4$ m/s. We get $I_y = 6 \times 10^{11}$ m/s. That is one gigantic specific impulse. No present **rocket engine** in the World has such a specific impulse and a rival is unlikely in the future.

For comparison the specific impulse are: conventional liquid-propellant rocket engine had maximum $I_y = 4200$ m/s; hydrogen rocket engine - $I_y = 5180$ m/s; thermonuclear rocket engine (H$_2$+H$_3$)(for 100% efficiency) has $I_y = 26 \times 10^6$ m/s; ideal laser engine has $I_y = 3 \times 10^8$ m/s; and the most power – annihilation rocket engine (for 100% efficiency) has theoretical impulse $I_y = 4.24 \times 10^8$ m/s.

Our AB-Space Engine has very high specific impulse and it may be a good candidate for interstellar flights. This system spends less mass for producing the plasma cable than any rocket engine spends for producing thrust. Another advantage is that it gets the energy from a (planet-mounted) station, i.e. the power source needn't travel with it and weigh it down. The AB-Engine is very light, simple, safety, and reliable with comparison to any likely (or perhaps nearly any dream) nuclear engine. In most cases at least part of the cable mass can be injected from the planet-mounted energy station.

*Coefficient efficiency of AB-Space Engines.* Author offers the following the estimation efficiency of AB-Space Engines: the ratio of energy (power) getting by apparatus to energy (power) spending by station:

$$\eta = \frac{PV}{N}, \qquad (28)$$

where $\eta$ is coefficient efficiency; $P$ is full thrust getting by apparatus, N; $V$ is apparatus speed, m/s; $N$ is electric station power, W. The formulas above allow to compute it, but one is variable value.

The other coefficient efficiency is the ratio the apparatus thrust to spending power of electric station [N/W]:

$$\eta_N = \frac{P}{N}, \quad P = \frac{\mu_0}{2\pi} I^2, \quad N = I^2 R, \quad R = \rho \frac{L}{s},$$

$$\rho = 1.03 \cdot 10^{-4} \ln \Lambda \cdot T^{-3/2}, \quad T = 2.71 \cdot 10^{-12} \left( \frac{j}{en} \right)^2 . \qquad (29)$$

Substitute all equations (29) in the first equation (29) we receive

$$\eta_N = 1.06 \cdot 10^{36} \frac{s}{L \cdot \ln \Lambda} \left( \frac{j}{n} \right)^3, \qquad (30)$$

*Estimation.* For $n = 10^{14}$ /m$^3$, $j = 10^3$ A/m$^2$ , $s = 10^2$ m$^2$, $L = 10^5$ m, ln$\Lambda = 10$ we get $\eta_N = 0.1$ N/W. However for high $L$ the coefficient is very small. That is because the electric station spends a lot of energy for producing magnetic field of closed-loop cables.

# Project

 As a example, we estimate the parameters of the AB-Space Engine having the thrust from one plasma cable about $P = 10^5$ N = 10 tons. Our design is not optimal. That is only simple of calculation.

Most our computation is made for one cable. For reality (two cable) engine you must double all values.

 Let us take the following initial data: electric current $I = 10^6$ A, thrust of one cable $P = 10^5$ N = 10 tons, radius of plasma cable $r = 10$ m, plasma density $n = 10^{14}$/m$^3$. $I > I_{\min} = 13$ A.

 We get the following results:

1) Density of electric current and electron speed are

$$j = \frac{I}{\pi r^2} = \frac{10^6}{3.14 \cdot 10^2} = 3180 \quad A/m^2, \quad u \approx \frac{j}{en} = \frac{3180}{1.6 \cdot 10^{-19} 10^{14}} = 2 \cdot 10^8 \quad m/s,$$

2) Temperature electrons and ions along cable axis and Spitzer electric resistance (for ln$\Lambda$ = 10) are

$$T = 2.71 \cdot 10^{-12} u^2 = 1.08 \cdot 10^5 \quad eV, \quad \rho = 1.03 \cdot 10^{-4} \ln \Lambda \cdot T^{-3/2} = 0.92 \cdot 10^{-10.5} \quad \Omega \cdot m.$$

3) Electric resistance, requested voltage and electric power for one plasma cable and its length $L = 10^{12}$ m = 1000 millions of km (Remain: distance from Sun to Earth is 150 millions of km, from Sun to Mars is 228 millions of km, from Sun to Jupiter 778 millions of km, and from Sun to Saturn 1427 millions of km).

$$R = \rho \frac{L}{s} = 0.92 \cdot 10^{-10.5} \frac{10^{12}}{3.14 \cdot 10^2} = 9.5 \cdot 10^{-2} \quad \Omega,$$

$$U = IR = 10^6 \cdot 9.5 \cdot 10^{-2} = 9.5 \cdot 10^4 \quad V, \quad N = IU = 10^6 \cdot 9.5 \cdot 10^4 = 9.5 \cdot 10^{10} \quad W.$$

The USA produced about 1022 GW electric energy in winter 2007. The 950 GigaWatts is about 20% power of all kinds of energy produced in the USA in 2005 or of electric power in the World. Our thrust is very high. If we take the thrust 1,000 – 10,000 N the requested electric power decreases in hundreds-thousands times. The apparatus can reach a needed speed by increasing the acceleration distance.

4) The mass of one cable is

$$M = sLnm_i = 3.14 \cdot 10^2 10^{12} 10^{14} 2 \cdot 1{,}67 \cdot 10^{-27} = 105 \quad kg$$

For two (forward and back) cables the cable mass is 210 kg. That is small mass for a cable having cross-section $s = 314$ m$^2$ (diameter 10 m) and length $L = 1000$ millions of km. That mass may be significantly less if we take the less plasma density. Part of this mass (about half) may be ejected from start electric station.

6) The magnetic thrust of cables (Fig.5g,h) located about apparatus at distance $d = 110$ m is

$$P = \frac{\mu_0 I^2}{2\pi} \ln \left| \left(\frac{d-r}{r}\right)^2 \right| = \frac{4\pi 10^{-7} 10^{12}}{2\pi} \ln \left| \left(\frac{110-10}{10}\right)^2 \right| = 9.2 \cdot 10^5 \quad N = 92 \, tons$$

 That thrust is different from initial thrust of two single plasma cables (20 tons) because we take the distance between plasma cable $d = 110$ m and receive ln 100 = 4.6 (not one).

7) The kinetic thrust of charged particles of one plasma cable is

$$P_e = 3.36 \cdot 10^7 \frac{j^2}{n} s = 3{,}36 \cdot 10^7 \frac{I^2}{sn} = 3.36 \cdot 10^7 \frac{10^{12}}{314 \cdot 10^{14}} = 1.07 \cdot 10^3 \quad N \approx 107 \, kg$$

 As you see the kinetic (electric) thrust is small (in comparison with magnetic thrust). We can neglect it.

 The total kinetic energy (charged particles) of one cable the length $L = 1000$ millions of km is

$$E = P_e L = 1.07 \cdot 10^3 10^{12} = 1.07 \cdot 10^{15} \quad J.$$

8) The additional voltage between the two cable line for balance magnetic (repeal) and electric (attracted) force and a power which must use for this the space ship or electric station are

$$\Delta U = 60 \cdot I \cdot \ln\left(\frac{d}{r}\right) = 60 \cdot 10^6 \cdot \ln 10 = 1.38 \cdot 10^8 \ V, \quad \Delta P = \Delta U \cdot I = 1.38 \cdot 10^8 \cdot 10^6 = 1.38 \cdot 10^{14} \ W.$$

That is large power. It may be requested the lower current and thrust or use the fig.5h design.

9) Estimation of flight possibilities the space ship having mass $M_s$ = 92 tons, flight time 10 days = $t = 8.64 \times 10^5$ sec, and computed AB-Engine. The acceleration of the space ship, speed and range are:

$$a = \frac{P}{M_s} = \frac{9.2 \cdot 10^5}{9.2 \cdot 10^4} = 10 \ m/s. \quad V = at = 10 \cdot 8.64 \cdot 10^5 = 8.64 \cdot 10^6 \ m/s = 8.64 \cdot 10^3 \ km/s,$$

$$L_s = \frac{at^2}{2} = \frac{10 \cdot (8.64 \cdot 10^5)^2}{2} = 3.73 \cdot 10^{12} \ m = 3730 \ millions \ of \ km.$$

This is in 50 times more then minimal distance from Earth to Mars.

Our estimation is not optimal, that is example of computation.

Note that the $V$ displayed is something on the order of $2.88\%$ of lightspeed!

## Discussion

*Advantages of AB-engine:*

1. The offered AB-Space Engine is very light, simple, safe, and reliable with comparison to any likely nuclear engine.
2. The AB-Space Engine has a gigantic 'virtual specific impulse', being more capable of realistic operation in a projectable near-future environment, than virtually any proposed means of thermonuclear or light-propulsion scheme the author is aware of.
3. The AB-Space Engine can accelerate a near-term space apparatus to very high speed (approaching light speed). At present time this is the single real method to be able to approach this 'ultimate' velocity.
4. At least part of the needed injected plasma cable mass and nearly all of the energy needed (and the cooling facilities needed to maintain that energy supply) can be from the planet-bound energy-supplying station, further improving the on-board 'mass ratio'.
5. The AB-Space Engine can use far cheaper energy from a planet-bound electric station.

The offered ideas and innovations may create a jump in space and energy industries. Author has made initial base researches that conclusively show the big possibilities offered by the methods and installations proposed. Further research and testing are necessary. Those tests are not expensive. As that is in science, the obstacles can slow, even stop, applications of these revolutionary innovations. For example, the plasma cable may be unstable. The instability mega-problem of a plasma cable was found in tokomak R&D, but it is successfully solved at the present time. The same method (rotation of plasma cable) can be applied in our case.

The other problem is production of the plasma cable in Earth's atmosphere. This problem may be sidestepped by operations from a suitably high super-stratospheric tower such as outlined in others of the author's works, or is no problem at all if the electric station of the plasma cables' origin is located on the Moon [8].

The author has ideas on how to solve this problem with today's technologies and to use the readily available electric stations found on this planet Earth. Inquiries from serious parties are invited.

## Summary

This new revolutionary idea – The AB-Space Engine and wireless transferring of electric energy in the hard vacuum of outer space – is offered and researched. A rarefied plasma power cord in the function of electric cable (wire) is used for it. It is shown that a certain minimal electric current creates a compression force that supports and maintains the plasma cable in its compacted form. Large amounts of energy can be transferred hundreds of millions of kilometers by this method. The requisite mass of plasma cable is merely hundreds of grams (some kg). A sample macroproject is computed: An AB-Space Engine having thrust = 10 tons. It is also shown that electric current in plasma cord can accelerate or slow various kinds of outer space apparatus.

## Acknowledgement

The author wishes to acknowledge Joseph Friedlander for correcting the author's English and useful technical advice and suggestions.

## References

(Reader finds some of author's articles in http://Bolonkin.narod.ru/p65.htm, http://arxiv.org , search "Bolonkin" and in books: "Non-Rocket Space Launch and Flight", Elsevier, 2006, 488 ps; and "New Concepts, Ideas, and Innovations in Aerospace, Technology and Human Life", NOVA, 2008, 400 ps.)